\documentclass[aps, twocolumn, prd, 10pt, showkeys, showpacs]{revtex4-1}
\usepackage{amssymb}
\usepackage{amsmath}
\usepackage[applemac]{inputenc}

\begin{document}
\title{Area products for stationary black hole horizons}
\author{Matt Visser}
\email{matt.visser@msor.vuw.ac.nz}
\affiliation{School of Mathematics, Statistics, and Operations Research, \\
Victoria University of Wellington, PO Box 600, Wellington 6140, New Zealand }
\date{6 June 2012; 24 July 2013; \LaTeX-ed \today}
\begin{abstract}
Area products for multi-horizon stationary black holes often have intriguing properties, and are often (though not always) independent of the mass of the black hole itself (depending only on various charges, angular momenta,  and moduli). Such products are often formulated in terms of the areas of inner (Cauchy) horizons and outer (event) horizons, and sometimes include the effects of unphysical ``virtual'' horizons.  But the conjectured mass-independence sometimes fails. Specifically, for the Schwarzschild--de~Sitter [Kottler] black hole in (3+1) dimensions  it is shown by explicit exact calculation that the product of event horizon area and cosmological horizon area is \emph{not} mass independent. (Including the effect of the third ``virtual'' horizon does not improve the situation.) Similarly, in the Reissner--Nordstrom-anti-de~Sitter black hole in (3+1) dimensions the product of inner (Cauchy) horizon area and event horizon area is calculated (perturbatively), and is shown to be \emph{not} mass independent. That is, the mass-independence of the product of physical horizon areas is \emph{not} generic. In spherical symmetry, whenever the quasi-local mass $m(r)$ is a Laurent polynomial in aerial radius, $r=\sqrt{A/4\pi}$, there are significantly more complicated mass-independent quantities, the elementary symmetric polynomials built up from the complete set of horizon radii (physical \emph{and} virtual). Sometimes it is possible to eliminate the unphysical virtual horizons, constructing combinations of physical horizon areas that are mass independent, but they tend to be considerably more complicated than the simple products and related constructions currently being mooted in the literature.
\end{abstract}
\keywords{horizon areas; event horizons; Cauchy horizons; cosmological horizons; 1205.6814 [hep-th].\\
Physical Review D {\bf88} (2013) 044014; doi: 10.1103/PhysRevD.88.044014}
\pacs{04.70.-s; 04.70.Bw; 04.70.Dy; 04.20.Jb}

\maketitle
\def\R{{\mathbb{R}}}
\def\N{{\mathbb{N}}}
\def\Z{{\mathbb{Z}}}
\def\Q{{\mathbb{Q}}}
\def\O{{\mathcal{O}}}

\clearpage
\section{Introduction}
\label{S:intro}

There has recently been some considerable ongoing interest in the products of horizon areas for various types of stationary black hole. Some of this interest has arisen specifically within the general relativity community~\cite{Ansorg:2008, Ansorg:2009, Hennig:2009, Ansorg:2010}, while for somewhat different reasons interest has also arisen from within the string community~\cite{Cvetic:2010, Castro:2012, Cvetic:2013, Larsen:2013}. In some cases the product of horizon areas is in fact independent of the mass of the black hole. 

For instance, based on classical general relativistic techniques it is known that both for standard (3+1) dimensional Kerr--Newman, and even for (3+1) dimensional Kerr--Newman black holes distorted by the presence of arbitrary stationary axisymmetric matter, the product of inner [Cauchy] horizon area and outer [event] horizon areas is~\cite{Ansorg:2008, Ansorg:2009, Hennig:2009, Ansorg:2010}:
\begin{equation}
A_C \;A_E  = (8\pi)^2 \left[ J^2  + {Q^4\over4} \right].
\label{E:product}
\end{equation}
The underlying physics here is that due to stationarity there can be no matter present between the inner and outer horizons (where the “radial” direction is timelike)~\cite{private}. The region between inner and outer horizons is then stationary, axisymmetric, and electro-vac;  this is not quite enough to be able to apply the black hole uniqueness theorems, but it appears that enough of the flavour of uniqueness survives to guarantee that the area product is not only independent of the mass of the black hole, but more remarkably is independent of the way the static axisymmetric matter external to the black hole (and distorting its gravitational field away from exact Kerr--Newman) is distributed.  These results are closely related to mass-independent inequalities for the area of the outer Killing horizon in stationary axisymmetric black holes with surrounding matter~\cite{Hennig:2008, Jaramillo:2011, Dain:2011, Simon:2011, Dain:2011b, Clement:2011, Jaramillo:2012, Clement:2012, Jaramillo:2012b, Dain:2013, Dain:GR20}:
\begin{equation}
A_E  \geq 8\pi \sqrt{J^2  + {Q^4\over4}}.
\label{E:inequality}
\end{equation}

Apart from the standard (3+1) dimensional Kerr--Newman spacetime, there are also many multi-dimensional string-inspired black hole configurations for which similar formulae hold~\cite{Cvetic:2010, Castro:2012, Cvetic:2013, Larsen:2013}. 
More boldly, there are also conjectures to the effect that this product of areas is sometimes quantized. 
That is, in the supersymmetric extremal limit one often finds
\begin{equation}
A_C \;A_E  = (8\pi)^2 \; L_P^4 \; N \quad\hbox{with}\quad N \in \N.
\end{equation}
For specific discussion of potential pitfalls for such a conjecture see \cite{Visser:2012, Faraoni:2012}. 
A safer statement is that when one moves away from extremality and supersymmetry then quite often the product of areas is discretized in terms of the Planck area and fine structure constant with
\begin{equation}
A_C \;A_E  = (8\pi)^2 L_P^4\left\{\ell(\ell+1)+{\alpha^2 q^2\over4}\right\},
\end{equation}
or some natural generalization thereof~\cite{Visser:2012}.
Here $\ell\in\N$ and $q\in \Z$.

But how generic are such mass-independence results? For instance, to what extent do they survive introduction of a  cosmological constant? It is already known that the area inequalities behave in a more complicated manner once a cosmological constant is introduced~\cite{Simon:2011, Clement:2012}.  Herein we address this issue in an elementary way by straightforwardly exhibiting several simple spherically symmetric (3+1) dimensional examples where, due precisely to a non-zero cosmological constant, the product of physical horizon areas is explicitly \emph{not} mass independent.  We shall explicitly consider the Schwarzschild-(anti)-de~Sitter and Reissner--Nordstr\"om-(anti)-de~Sitter spacetimes, before considering general lessons we can extract for generic static spherically symmetric spacetimes.
The fact that asymptotically anti-de~Sitter black holes often fail to have mass-independent area products is perhaps of most interest  
to the string community, indicating that more complicated functions of horizon area might be of interest.

There will typically be \emph{some} (sometimes several) more complicated functions of physical horizon areas that are mass independent, but generically these functions are nowhere near as straightforward as a simple product of areas.
As we shall soon see, obtaining mass independent functions of horizon areas in spherical symmetry is intimately related to the quasi-local mass $m(r)$ being a Laurent polynomial of the areal radius $r$ defined by $A(r) =4\pi r^2$. (Because of spherical symmetry the quasi-local mass is always guaranteed to be well defined, and so is a sufficiently general tool for the current article. Any attempt at moving to axisymmetry would require slightly more subtle tools; the norm of the horizon-generating Killing vector is an appropriate quantity to consider.)  The relevant mass-independent area-related functions are constructed in terms of the elementary symmetric polynomials built up from the radii of the various horizons (both physical \emph{and} virtual).  Sometimes one can eliminate the virtual horizons to obtain more complicated mass-independent qualities depending only on the physical horizons.   

\section{Framework}
\label{S:framework}

Based only on symmetry one can without any loss of generality write any static spherically symmetric spacetime in the form~\cite{dirty}
\begin{eqnarray}
ds^2 &=& - \exp\{2\Phi(r)\} \left(1-{2 m(r)\over r} \right) dt^2 
\nonumber\\
&&
+ {dr^2\over 1 - 2m(r)/r} + r^2 \{ d\theta^2 + \sin^2\theta\,d\varphi^2\}. 
\end{eqnarray}
Here $m(r)$ denotes the quasi-local mass~\cite{hernandez-misner, misner-sharp}, and $\Phi(r)$ is the anomalous redshift~\cite{dirty}. 
The Killing horizons are then found by solving
\begin{equation}
\Delta(r) \equiv 1-{2 m(r)\over r} = 0.
\end{equation}
Once we have extracted the various roots of this equation, the individual horizon areas are immediate.

\section{Schwarzschild--de~Sitter black holes}
\label{S:S-dS}
For Schwarzschild--de~Sitter [Kottler] black holes the Killing horizons are found by solving the equation
\begin{equation}
\Delta(r)= 1-{2m\over r}-{1\over3} \Lambda r^2 = 0.
\end{equation}
This is equivalent to solving the cubic
\begin{equation}
r^3 -3 r/\Lambda + 6m/\Lambda=0.
\end{equation}
For $\Lambda>0$ it is convenient to set $\Lambda=1/a^2$, where $a$ is (asymptotically) the spatial radius of curvature. Then
\begin{equation}
r^3 - 3 r a^2 + 6m a^2 = 0.
\end{equation}
The three \emph{exact} roots for this cubic are (see appendix \ref{S:cubic})
\begin{equation}
r= 2a \sin\left( {1\over3} \sin^{-1}\left({3m\over  a} \right) + \epsilon {2\pi\over3}\right); \quad \epsilon\in\{0,\pm1\}.
\end{equation}

\subsection{Killing horizons}
The two physical roots are the event horizon at
\begin{eqnarray}
r_E &=& 2a \sin\left({1\over3}\sin^{-1}\left({3m\over  a} \right)  \right) 
\nonumber\\
&=& 2m +{8m^3\over 3 a^3} +\O\left({m^5\over a^4}\right)  ,
\end{eqnarray}
and the cosmological horizon at
\begin{eqnarray}
r_\Lambda&=&
2a \sin\left({2\pi\over 3} + {1\over3}\sin^{-1}\left({3m\over  a} \right) \right) 
\nonumber\\
&=&  \sqrt{3} a - m +\O\left({m^2\over a}\right).
\end{eqnarray}
There is a third (unphysical and purely formal) ``virtual'' horizon which is located at negative $r$:
\begin{equation}
\label{E:SdS1}
r_V = - r_\Lambda- r_E. 
\end{equation}
Note that the product of physical horizon areas, $A_E\times A_\Lambda$, has no nice quantization features. Nor does it have any nice ``independence of mass'' features. Indeed
\begin{eqnarray}
A_E \times A_\Lambda &=&
 (16\pi a^2)^2 \sin^2\left({2\pi\over 3} + {1\over3}\sin^{-1}\left({3m\over  a} \right) \right)
 \nonumber\\
 && \qquad\times\sin^2\left({1\over3}\sin^{-1}\left({3m\over  a} \right)  \right)
\\
&=& (8\pi a^2)^2 \left[ \cos\left({2\pi\over 3} + {2\over3}\sin^{-1}\left({3m\over  a} \right) \right) -  {1\over2} \right]^2
\nonumber\\
&&
\\
&=& (8\pi)^2 \bigg\{ 3 m^2 a^2 -2\sqrt{3} m^3 a  + 6 m^4 
\nonumber\\
&&\qquad\qquad+\O(m^5/a)  \bigg\}.
\end{eqnarray}
If one restricts attention to the two physical horizons at the two physical roots of the cubic,  then  in terms of area products this is the best one can do. If one includes the effect of the virtual horizon $r_V$, as advocated in reference~\cite{Cvetic:2010},  then we have the exact results
\begin{equation}
r_V \; r_E \; r_\Lambda = - 6 m a^2; \qquad A_V \; A_E \; A_\Lambda =  (4\pi)^3 \; 36 m^2 a^4.
\end{equation}
These are, however,  explicitly mass dependent quantities.
%

\subsection{Mass independence}
In counterpoint, note that there \emph{is} an exact mass-independent quantity arising from a \emph{quadratic} sum over all three roots of the cubic. Namely:
\begin{equation}
\sum_{i>j} r_i r_j = - 3 a^2.
\end{equation}
That is
\begin{equation}
\label{E:SdS2}
r_V \{ r_E+ r_\Lambda \} +  r_E\, r_\Lambda = - 3 a^2.
\end{equation}
We can eliminate the “virtual” radius and rewrite this as
\begin{equation}
 \{ r_\Lambda+ r_E \}^2 -  r_\Lambda r_E =  3 a^2,
\end{equation}
and so
\begin{equation}
\label{E:SdS3a}
 r_\Lambda^2 + r_E^2 +  r_\Lambda  r_E =  3 a^2.
\end{equation}
If one prefers to work in terms of areas one has
\begin{equation}
\label{E:SdS3b}
 A_\Lambda + A_E +  \sqrt{A_\Lambda  A_E} =  12 \pi a^2.
\end{equation}
So there is certainly \emph{some} function of physical horizon areas that is mass independent, but the function that exhibits mass independence is nowhere near as straightforward as a simple product of horizon areas.  

\section{Schwarzschild-anti-de~Sitter black holes}
\label{S:S-adS}
Consider the Schwarz\-schild-anti-de~Sitter black hole. Now set $\Lambda = - 1/|a|^2$. We determine the Killing horizons via the polynomial
\begin{equation}
r^3 + 3 r |a|^2 - 6m |a|^2 = 0.
\end{equation}
There is now only one physical root, only one physical horizon (an event horizon), located at
\begin{equation}
r_E= 2|a| \sinh\left( {1\over3} \sinh^{-1}\left({3m\over  |a|} \right)\right).
\end{equation}
To make this fully explicit, in terms of Cardano's formulae one can rewrite this as 
\begin{eqnarray}
r_E &=& |a| \Bigg\{ 
\left[ \sqrt{1+{9m^2\over|a|^2}} + {3m\over|a|} \right]^{1/3} 
\nonumber\\
&&\qquad
-  
\left[  \sqrt{1+{9m^2\over|a|^2}} - {3m\over|a|}\right]^{1/3}   
\Bigg\}.
\end{eqnarray}
There are now two purely formal and unphysical virtual horizons, at  complex conjugate values of $r_V$ and $r _V^*$. There is an exact result that
\begin{equation}
r_V \; r_V^* \; r_E =  6 m |a|^2; \quad A_V \; A_V^* \; A_E =  (4\pi)^3 \; 36 m^2 a^4.
\end{equation}
This is again explicitly mass dependent. The mass-independent quantity constructed from the event horizon and two virtual horizons is now
\begin{equation}
\sum_{i>j} r_i r_j =  3 |a|^2.
\end{equation}
That is
\begin{equation}
r_E \{ r_V+ r_V^* \} +  r_V\, r_V^* = 3 |a|^2. 
\end{equation}
We can simplify this a little by noting that
\begin{equation}
\label{E:SadS1}
r_E + r_V + r_V^*  = 0,
\end{equation}
so that
\begin{equation}
\label{E:SadS2}
r_E^2 = r_V r_V^* - 3 |a|^2;  \qquad A_E = |A_V| - 12\pi |a|^2.
\end{equation}
This is at least formally mass independent --- but since $|A_V|$ is not directly observable,  (and not calculable except by explicitly solving the mass-dependent cubic), the result is not particularly useful.

\section{Reissner--Nordstr\"om--de~Sitter black holes}
\label{S:RN-dS}
The situation improves somewhat for Reissner--Nordstr\"om--de~Sitter black holes. 
To locate the Killing horizons  we need to find the roots of
\begin{equation}
\Delta(r)= 1-{2m\over r} + {Q^2\over r^2}-{1\over3} \Lambda r^2 =0.
\end{equation}
Again setting $\Lambda=1/a^2$, we now rearrange this to obtain the quartic
\begin{equation}
r^4 - 3 r^2 a^2 + 6m r a^2 - 3 Q^2 a^2= 0.
\end{equation}
Taking $\Lambda\to0$, (corresponding to $a\to\infty$), gives the standard Reissner--Nordstr\"om geometry.  Also, $Q\to0$ gives the Schwarzschild--de~Sitter [Kottler] solution previously considered. 
Let us now write the quartic as 
\begin{equation}
r^4 - 3 a^2 \{ r^2 - 2m r + Q^2\} = 0,
\end{equation}
and reformulate  this as
\begin{equation}
r^4 -3 a^2 (r-r_+)(r-r_-)=0.
\end{equation}
Here $r_\pm$ are the locations where the horizons would be in the limit where the cosmological constant is switched off ($\Lambda\to0$, that is, $a\to\infty$).  For simplicity we shall take $|Q|\leq m$, so that the $r_\pm$ are guaranteed real. (There is no real point to considering the sub-case where $r_\pm$ are complex.)

\subsection{Approximate results}
While we know on general principles that the quartic appearing above has an exact solution, it can be more advantageous to perturbatively extract approximate solutions. 
First, rearrange the quartic to yield the exact equation
\begin{equation}
r = r_\pm + {r^4\over 3 a^2(r-r_\mp)}.
\end{equation}
We shall now solve this equation perturbatively.

\subsubsection{Event and Cauchy horizons}
To a first approximation, for the event horizon we have
\begin{equation}
r_E \approx r_+ + {r_+^4\over 3 a^2(r_+-r_-)} = r_+ \left\{ 1 + {r_+^3\over 3 a^2(r_+-r_-)} \right\}.
\end{equation}
For the inner (Cauchy) horizon we have
\begin{equation}
r_C \approx  r_- - {r_-^4\over 3 a^2(r_+-r_-)} = r_- \left\{ 1 - {r_-^3\over 3 a^2(r_+-r_-)}  \right\}.
\end{equation}
Consequently
\begin{equation}
r_E \; r_C \approx r_+\; r_- \left\{ 1 + {r_+^3-r_-^3\over 3 a^2(r_+-r_-)} \right\},
\end{equation}
and so
\begin{equation}
r_E \; r_C \approx r_+\; r_- \left\{ 1 + {r_+^2+   r_+ r_-   +r_-^2\over 3 a^2} \right\}.
\end{equation}
But in terms of the mass and charge we know
\begin{equation}
r_\pm = m \pm \sqrt{m^2-Q^2},
\end{equation}
whence
\begin{equation}
r_+\, r_- = Q^2,
\end{equation}
and
\begin{equation}
r_\pm^2 = 2m^2-Q^2 \pm 2m\sqrt{m^2-Q^2},
\end{equation}
so
\begin{equation}
r_+^2+ r_+ r_- +r_-^2 = 4m^2 - Q^2.
\end{equation}
This implies
\begin{equation}
r_E \; r_C \approx Q^2 \left\{ 1 + {4m^2-Q^2\over 3 a^2} \right\},
\end{equation}
which can also be written as
\begin{equation}
r_E\; r_C = Q^2 \left\{ 1 + {1\over3}\Lambda(4m^2-Q^2) + \mathcal{O}(\Lambda^2) \right\}.
\end{equation}
Therefore
\begin{equation}
A_E \; A_C= 16\pi^2 Q^4 \left\{ 1 + {2\over3}\Lambda(4m^2-Q^2) + \mathcal{O}(\Lambda^2) \right\},
\end{equation}
which is certainly \emph{not} mass independent. 

For completeness we also note
\begin{equation}
r_E + r_C \approx 2m + {r_+^4-r_-^4\over 3 a^2(r_+-r_-)} ,
\end{equation}
which again is explicitly mass dependent. 

\subsubsection{Cosmological horizon}
From the exact result
\begin{equation}
r^2 = 3 a^2 \; {(r-r_+)(r-r_-)\over r^2},
\end{equation}
we have, as a zero order approximation,
\begin{equation}
r_\Lambda \approx \sqrt{3} a.
\end{equation}
Therefore as a first order approximation 
\begin{eqnarray}
r_\Lambda &\approx& \sqrt{3} a \sqrt{ (\sqrt{3} a-r_+)(\sqrt{3} a-r_-)\over 3 a^2} 
\\
&\approx& \sqrt{3} a \left\{ 1 - {r_++r_-\over 2\sqrt{3} a} \right\}
\\
&=&
\sqrt{3} a \left\{ 1 - {m\over \sqrt{3} a} \right\}.
\end{eqnarray}
So for the cosmological horizon
\begin{equation}
r_\Lambda \approx \sqrt{3} a - m.
\end{equation}
Oddly enough the location of the cosmological horizon is to this order independent of the charge $Q$, but it does definitely depend on the mass $m$. 

\subsubsection{Virtual horizon}
Finally, from the exact quartic, we know there is a (unphysical) virtual horizon at negative $r$:
\begin{equation}
\label{E:RNdS1}
r_V = -  \{ r_E + r_C + r_\Lambda \}.
\end{equation}
So to a first approximation
\begin{equation}
r_V \approx - \sqrt{3} a - m. 
\end{equation}

\subsection{Exact results}
What quantities might actually be independent of $m$? From the exact quartic we know
\begin{equation}
\label{E:RNdS2}
r_V \; r_E \; r_C \;  r_\Lambda   =  -3 Q^2 a^2,
\end{equation}
implying, in terms of physical horizons, that the quantity
\begin{equation}
\label{E:RNdS4}
 \{ r_E + r_C + r_\Lambda \} \; r_E\; r_C\; r_\Lambda   =  3 Q^2 a^2 
\end{equation}
is strictly independent of $m$. But this looks nothing like the product of event and Cauchy horizon areas $A_+ A_-$. 
Perhaps more promising is the exact condition
\begin{equation}
\sum_{i>j} r_i r_j = - 3 a^2. 
\end{equation}
That is
\begin{equation}
\label{E:RNdS3}
r_V \,\{r_E  + r_C +  r_\Lambda \}
+r_E \, \{ r_C  + r_\Lambda \} +  r_C  \,r_\Lambda 
= - 3 a^2,
\end{equation}
whence
\begin{equation}
\{r_E  + r_C +  r_\Lambda \}^2
-r_E  \,\{ r_C  + r_\Lambda \} -  r_C  \, r_\Lambda 
= 3 a^2,
\end{equation}
so that
\begin{equation}
\label{E:RNdS5}
r_E^2 + r_C^2 +  r_\Lambda^2
+ r_E \, r_C  +r_C \, r_\Lambda +
 r_\Lambda \, r_E
= 3 a^2.
\end{equation}
We can furthermore eliminate explicit (though not implicit) occurrence of the cosmological constant by dividing these two exact results to get
\begin{equation}
\label{E:RNdS6}
{
\{ r_E + r_C + r_\Lambda \} \,r_E\,  r_C \, r_\Lambda
\over
r_E^2 + r_C^2 +  r_\Lambda^2
+ r_E  r_C  +r_C  \, r_\Lambda +
 r_\Lambda \,r_E
 }
 = Q^2.
\end{equation}
This is certainly mass independent, but is a rather complicated function of physical horizon radii.  
As $a\to\infty$ (that is $\Lambda\to 0$, so $r_\Lambda\to\infty$) one recovers the usual Reissner--Nordstr\"om result
\begin{equation}
\lim_{a\to\infty}  r_E  \, r_C = Q^2. 
\end{equation}
If one insists on working with areas then we have the exact result that $4 \pi Q^2$ is equal to 
\begin{eqnarray}
\label{E:RNdS7}
{
\{ \sqrt{A_E} + \sqrt{A_C} + \sqrt{A_\Lambda} \} \,\sqrt{A_E\,  A_C \, A_\Lambda}
\over
A_E^2 + A_C^2 +  A_\Lambda^2
+ \sqrt{A_E  A_C}  +\sqrt{A_C  \, A_\Lambda} + \sqrt{A_\Lambda \,A_E}
 }.
 \nonumber\\
 &&
\end{eqnarray}
Again, there is certainly \emph{some} function of physical horizon areas that is mass-independent, (and in this particular case, even free of explicit cosmological constant dependence), but it is nowhere near as straightforward as a simple product of horizon areas. 

\section{Reissner--Nordstr\"om-anti-de~Sitter black holes}
\label{S:RN-adS}

Set $\Lambda = -1/|a|^2$. The relevant quartic becomes
\begin{equation}
r^4 + 3 r^2 |a|^2 - 6m r |a|^2 + 3 Q^2 |a|^2= 0.
\end{equation}
There are now two complex conjugate (utterly formal and unphysical) virtual horizons $r_V^\pm$, and two physical horizons: an event horizon $r_E$ and an inner (Cauchy) horizon $r_C$.  Because there are only two physical horizons, this particular situation is closest in spirit to the standard Reissner--Nordstr\"om spacetime. 

\subsection{Approximate results}
To a first approximation, for the event horizon we have
\begin{equation}
r_E \approx r_+ - {r_+^4\over 3 |a|^2(r_+-r_-)} = r_+ \left\{ 1 - {r_+^3\over 3 |a|^2(r_+-r_-)} \right\}.
\end{equation}
For the inner [Cauchy] horizon we see
\begin{equation}
r_C \approx  r_-   + {r_-^4\over 3 |a|^2(r_+-r_-)} = r_- \left\{ 1 + {r_-^3\over 3 |a|^2(r_+-r_-)}  \right\}.
\end{equation}
Finally, for the two unphysical virtual horizons we obtain
\begin{equation}
r_V^\pm \approx \pm i \sqrt{3} \, |a|  - m.
\end{equation}
Then it is easy to compute
\begin{equation}
r_E \,r_C \approx r_+ \,r_- \left\{ 1 - {r_+^3-r_-^3\over 3 |a|^2(r_+-r_-)} \right\},
\end{equation}
so that 
\begin{equation}
r_E \,r_C \approx r_+ \,r_- \left\{ 1 - {r_+^2+ r_+ r_-r_+ +r_-^2\over 3 |a|^2} \right\},
\end{equation}
and so 
\begin{equation}
r_E \, r_C \approx Q^2 \left\{ 1 - {4m^2-Q^2\over 3 |a|^2} \right\}.
\end{equation}
Then,  (and I again emphasize that  for $\Lambda<0$ we are in an asymptotically adS spacetime with no cosmological horizon, and we really only have these two physical horizons to deal with), we see
\begin{equation}
r_E \,r_C = Q^2 \left\{ 1 - {1\over3}|\Lambda|(4m^2-Q^2) + \mathcal{O}(\Lambda^2) \right\}.
\end{equation}
In fact this now implies  that for \emph{either} sign of the cosmological constant one has
\begin{equation}
r_E \,r_C = Q^2 \left\{ 1 +{1\over3}\Lambda(4m^2-Q^2) + \mathcal{O}(\Lambda^2) \right\}. 
\end{equation}
Note this is very definitely \emph{not} mass independent. 

\subsection{Exact results}
Some exact results can again be obtained by computing various combinations of the roots of the quartic. Note that the key basic results obtained by picking off the various coefficients of the quartic are:
\begin{equation}
\label{E:RNadS1}
r_V^+ + r_V^- + r_C + r_E = 0;
\end{equation}
\begin{equation}
\label{E:RNadS2}
r_V^+ \,r_V^- + (r_V^+ + r_V^-) \,(r_C + r_E) =  3|a|^2;
\end{equation}
\begin{equation}
r_V^+ \,r_V^- \,( r_C + r_E) + (r_V^+ + r_V^-)\, r_C\,  r_E   = - 6 m |a|^2;
\end{equation}
and
\begin{equation}
\label{E:RNadS3}
r_V^+  \,r_V^- \, r_C \, r_E = 3 Q^2 |a|^2. 
\end{equation}
Therefore
\begin{equation}
(r_V^+ + r_V^-) = - (r_C + r_E); \qquad r_V^\pm =  - {1\over2}(r_C + r_E) \pm i \gamma,
\end{equation}
and so
\begin{equation}
r_V^+ \, r_V^- =  3|a|^2 + (r_C + r_E)^2;
\end{equation}
\begin{equation}
(r_V^+ \,r_V^- - r_C  \, r_E)(r_C + r_E)   = - 6 m |a|^2;
\end{equation}
\begin{equation}
r_V^+ r_V^- = {1\over4}(r_C + r_E)^2 + \gamma^2.
\end{equation}
We can eliminate some of the unknowns in the above expressions but not all. \\
In particular
\begin{equation}
{r_V^+  r_V^-  r_C  r_E \over r_V^+ r_V^- + (r_V^+ + r_V^-) (r_C + r_E)} = Q^2,
\end{equation}
so
\begin{equation}
{[ {1\over4}(r_C + r_E)^2 + \gamma^2]  r_C  r_E \over \gamma^2 - {3\over4}(r_C + r_E)^2} = Q^2.
\end{equation}
Unfortunately, while the RHS depends only on the charge $Q$,  the LHS contains the parameter $\gamma$, which is not directly accessible to physical observation. (Nor is it easy to calculate without explicitly solving the quartic.)
Alternatively, one could also write
\begin{equation}
\label{E:RNadS4a}
\left\{3|a|^2 + (r_C + r_E)^2\right\} r_C  r_E = 3 Q^2 |a|^2. 
\end{equation}
Therefore
\begin{equation}
\label{E:RNadS4b}
\left\{1 + {1\over3} |\Lambda| (r_C + r_E)^2\right\} r_C  r_E =  Q^2. 
\end{equation}
This is at least $m$ independent, and $\gamma$ independent, but explicitly contains both $Q$ and $\Lambda$. 
If we work in terms of areas
\begin{equation}
\label{E:RNadS4c}
\left\{1 + {1\over12\pi} |\Lambda| \left(\sqrt{A_C} + \sqrt{A_E}\right)^2\right\} \sqrt{A_C  A_E} =  4\pi Q^2. 
\end{equation}
Again, there is \emph{some} function of the physical horizon areas that is mass-independent, but it is nowhere near as straightforward as a simple product of horizon areas.

\section{Laurent polynomial for the quasi-local mass}
\label{S:laurent}

Let us now try to put these specific results into a broader context. Suppose merely that the quasi-local mass $m(r)$ is some generic Laurent polynomial. Then without loss of generality $\Delta(r)$ is also a Laurent polynomial and can be written in the form 
\begin{equation}
\Delta(r) =  \Delta_* \; {P(r)\over r^n}. 
\end{equation}
Here we have normalized the (ordinary) polynomial $P(r)$ so that its highest degree coefficient is unity, and its lowest degree coefficient (a constant term) is nonzero. 
The Killing horizons are located at the zeros $r_i$ of the numerator $P(r)$. That is, we have
\begin{equation}
P(r) =  \sum_{j=0}^{D-1} c_j \; r^j  + r^D=  \prod_{i=1}^D ( r - r_i).
\end{equation}
Furthermore, as is completely standard:
\begin{equation}
c_0 =   (-1)^D \prod_{i=1}^D r_i; \qquad c_1 =  (-1)^{D-1} \sum_{j=1}^D  \prod_{i=1, i \neq j}^D r_i; \quad \dots
\end{equation}
\begin{equation}
\dots \quad c_{D-2} = \sum_{i>j} r_i r_j; \qquad c_{D-1} =  - \sum_{j=1}^D  r_j.
\end{equation}
In fact these coefficients are easily and explicitly calculable in terms of the elementary symmetric polynomials $e_i(\cdot)$ on $D$ variables~\cite{symmetric1, symmetric2}:
\begin{equation}
c_{D-i} = (-1)^D \; e_i(r_1,r_2,\dots, r_{D}). 
\end{equation}
We see that it is the coefficient $c_{n-1}$ that leads to a $1/r$ falloff in $\Delta(r)$ at large $r$, and so it is this coefficient that is proportional to the mass of the black hole. (By construction $n\in\{1,\dots,D\}$, otherwise the mass of the black hole will be zero.) All of the other coefficients, (there are $D-1$ of them),
\begin{equation}
c_i(r_1,r_2,\dots, r_{D}):  \quad   0 \leq i \leq D-1; \quad {i \neq n-1},
\end{equation}
will by construction be mass-independent. That is, in terms of the elementary symmetric polynomials, all the quantities
\begin{equation}
e_i(r_1,r_2,\dots, r_{D}):  \quad   1 \leq i \leq D; \quad {i \neq D-n+1},
\end{equation}
will be mass-independent. In terms of horizon areas, $A_i=4\pi r_i^2$, \emph{all} $D-1$ elementary symmetric polynomials
\begin{equation}
e_i\left( \sqrt{A_1\over4\pi},  \sqrt{A_2\over4\pi}, \dots,  \sqrt{A_D\over4\pi}\right):  \quad   1 \leq i \leq D, 
\end{equation}
for $ {i \neq D-n+1}$,
will be mass-independent. Of course not all the $r_i$ need be physical (real and positive), so not all the $A_i$ need be real. Since there are $D-1$ of these mass-independent quantities,  it might sometimes be possible to eliminate all the unphysical (virtual) horizons $r_i$,  and reduce the situation to one of dealing with a smaller number of real mass-independent quantities determined solely in terms of physical horizon areas.  With $N$ virtual horizons one will generally have $D-N-1$ mass independent quantities constructible in terms of physical horizons. Whether or not this can successfully be achieved in practice depends very much on the precise details of the polynomial $P(r)$. For example, as we have seen in the previous sections:
\begin{itemize}
\item Schwarzschild-de~Sitter spacetimes correspond to $D=3$ and $N=1$.
\\
There are two mass-independent quantities, [one trivial, equation~(\ref{E:SdS1}), one non-trivial, equation~(\ref{E:SdS2})],  but only one that depends solely on the physical horizons [equation~(\ref{E:SdS3a}) or equivalently~(\ref{E:SdS3b})].

\item Schwarzschild-anti-de~Sitter spacetimes correspond to $D=3$ and $N=2$. 
\\
There are two mass-independent quantities, [one trivial, equation~(\ref{E:SadS1}), one non-trivial, equation~(\ref{E:SadS2})], but none that depend solely on the physical horizon.

\item Reissner--Nordstr\"om--de~Sitter spacetimes correspond to $D=4$ and $N=1$. 
\\
There are three mass-independent quantities, [one trivial, equation~(\ref{E:RNdS1}), two non-trivial, equations~(\ref{E:RNdS2}) and (\ref{E:RNdS3})], but only two that depend solely on the physical horizons, [any two of equations~(\ref{E:RNdS4}), (\ref{E:RNdS5}), and (\ref{E:RNdS6}) --- or the equivalent (\ref{E:RNdS7})].

\item Reissner--Nordstr\"om-anti-de~Sitter spacetimes correspond to $D=4$ and $N=2$. 
\\
There are three mass-independent quantities, [one trivial, equation~(\ref{E:RNadS1}), two non-trivial, equations~(\ref{E:RNadS2}) and (\ref{E:RNadS3})], but only one that depends solely on the physical horizons, [any one of the equivalent equations~(\ref{E:RNadS4a}), (\ref{E:RNadS4b}), or (\ref{E:RNadS4c})].
\end{itemize}
But now we see that the key points of the preceding explicit discussion continue to hold in greater generality --- whenever the quasi-local mass $m(r)$ is any generic Laurent polynomial. Generalizations to higher dimensional spacetimes with hyper-spherical symmetry are immediate and straightforward. Generalizations to rotating black holes~\cite{Kerr:1963, Newman:1965a, Newman:1965b, Kerr:book, Kerr:survey},  and more complicated symmetries, are not quite as straightforward --- but as long as the location of the horizons is determined by the roots of some Laurent polynomial we can expect similar results to hold. For instance, it is quite sufficient if, in terms of some natural $r$ coordinate easily related to the horizon area, the norm of the horizon generating Killing vector is some entire function multiplied by a Laurent polynomial.

\section{Discussion}
\label{S:disc}

Generically, products of horizon areas may or may not be independent of the mass of the black hole. This depends on the precise form of the quasi-local mass, on whether one takes the product only over physical horizons, or whether one includes unphysical virtual horizons in the product.  In spherical symmetry, as long as the quasi-local mass is a Laurent polynomial with $D = D_\mathrm{max}-D_\mathrm{min}$,  there will be $D$ horizons from which one can construct $D-1$ mass-independent  quantities in terms of the elementary symmetric polynomials built out of the horizon radii. If $N$ of these horizons are ``virtual'' (negative or complex radius), then by algebraically eliminating the virtual horizons there will generally be $D-N-1$ (quite complicated) mass-independent  quantities constructible solely in terms of the physical horizon radii (and hence constructible in terms of the physical horizon areas). We have explicitly checked these results for validity by investigating the situation for Schwarzschild-(anti)-de~Sitter and Reissner--Nordstr\"om-(anti)-de~Sitter spacetimes. 

As we have seen above, with regard to string-inspired area products the general situation is much more complicated than currently envisaged. The conjectured area quantization generally fails because certain parameters are not integers~\cite{Visser:2012}. To quote the authors of~\cite{Horowitz:1996}: “we will refer to them as the numbers of branes, anti-branes, and strings because (as will be seen) they reduce to those numbers in certain limits where these concepts are well defined.” Furthermore, inspection of known exact solutions demonstrates that the conjectured mass independence often fails once a cosmological constant is added.

In contrast, for the general relativity inspired area bounds are not dependent on explicit exact solutions and at least partially survive the introduction of a cosmological constant~\cite{Simon:2011, Clement:2012}. There seems some hope of yet further progress along these lines. Similarly the Ansorg--Hennig--Cederbaum area product theorems~\cite{Ansorg:2008, Ansorg:2009, Hennig:2009, Ansorg:2010} are not dependent on explicit exact solutions  --- both the underlying framework and motivation is rather different --- as are the required tools.

\section*{Acknowledgments} 

This research was supported by the Marsden Fund, and by a James Cook Fellowship, both 
administered by the Royal Society of New Zealand.  

\clearpage
\appendix*
\section{Cubic polynomial equations}
\label{S:cubic}

Consider a cubic polynomial equation in reduced form, with coefficients conveniently chosen to be
\begin{equation}
x^3 -3 p^2x +2 q = 0;   \qquad p > 0.
\end{equation}
Then the exact roots are given by a form of Vi\`ete's trigonometric solution %
\begin{equation}
x = 
2 p \; \sin\left\{{1\over3}\;\sin^{-1}\left[{ q\over p^3} \right] + \epsilon \; {2\pi\over3}\right\}; \qquad \epsilon\in\{0,\pm1\}.
\end{equation}
If $|q|<p^3$ there are three real roots.  

On the other hand, if we have
\begin{equation}
x^3 +3 p^2x -2 q = 0;   \qquad p > 0,
\end{equation}
then there is only one real root. It is given by a hyperbolic form of Vi\`ete's  solution
\begin{equation}
x = 
2 p \; \sinh\left\{{1\over3}\sinh^{-1}\left[{ q\over p^3} \right] \right\}.
\end{equation}
In terms of Cardano's formulae one can explicitly rewrite this as 
\begin{equation}
x = p \left\{ 
\left[ \sqrt{1+{q^2\over p^6}} + {q\over p^3} \right]^{1/3} 
-  
\left[  \sqrt{1+{q^2\over p^6}} - {q\over p^3}\right]^{1/3}   
\right\}.
\end{equation}



\begin{thebibliography}{99}


  
  
  \bibitem{Ansorg:2008}
  M.~Ansorg and J.~Hennig,
  ``The Inner Cauchy horizon of axisymmetric and stationary black holes with surrounding matter'',
  Class.\ Quant.\ Grav.\  {\bf 25} (2008) 222001
  [arXiv:0810.3998 [gr-qc]].
  
   \bibitem{Ansorg:2009}
  M.~Ansorg and J.~Hennig,
  ``The Inner Cauchy horizon of axisymmetric and stationary black holes with surrounding matter in Einstein-Maxwell theory'',
  Phys.\ Rev.\ Lett.\  {\bf 102} (2009) 221102
  [arXiv:0903.5405 [gr-qc]].
  
  \bibitem{Hennig:2009}
  J.~Hennig and M.~Ansorg,
  ``The Inner Cauchy horizon of axisymmetric and stationary black holes with surrounding matter in Einstein-Maxwell theory: Study in terms of soliton methods'',
  Annales Henri Poincare {\bf 10} (2009) 1075
  [arXiv:0904.2071 [gr-qc]].
  
  \bibitem{Ansorg:2010}
  M.~Ansorg, J.~Hennig, and C.~Cederbaum,
  ``Universal properties of distorted Kerr-Newman black holes'',
  Gen.\ Rel.\ Grav.\  {\bf 43} (2011) 1205
  [arXiv:1005.3128 [gr-qc]].
  

\bibitem{Cvetic:2010}
  M.~Cvetic, G.~W.~Gibbons, and C.~N.~Pope,
  ``Universal Area Product Formulae for Rotating and Charged Black Holes in Four and Higher Dimensions'',
  Phys.\ Rev.\ Lett.\  {\bf 106} (2011) 121301
  [arXiv:1011.0008 [hep-th]].
  
\bibitem{Castro:2012}
  A.~Castro and M.~J.~Rodriguez,
  ``Universal properties and the first law of black hole inner mechanics'',
  Phys.\ Rev.\ D {\bf 86} (2012) 024008
  [arXiv:1204.1284 [hep-th]].
 
  
  \bibitem{Cvetic:2013}
  M.~Cvetic, H.~Lu and C.~N.~Pope,
  ``Entropy-Product Rules for Charged Rotating Black Holes'',
  arXiv:1306.4522 [hep-th].
 
\bibitem{Larsen:2013} 
Finn Larsen, “A Quantization Rule for Black Hole Horizons”, GR20, Warsaw, July 2013.


\bibitem{private}
Joerg Hennig, private communication.
  
  
\bibitem{Hennig:2008}
  J.~Hennig, M.~Ansorg and C.~Cederbaum,
  ``A Universal inequality between angular momentum and horizon area for axisymmetric and stationary black holes with surrounding matter'',
  Class.\ Quant.\ Grav.\  {\bf 25} (2008) 162002
  [arXiv:0805.4320 [gr-qc]].

\bibitem{Jaramillo:2011}
  J.~L.~Jaramillo, M.~Reiris and S.~Dain,
  ``Black hole Area-Angular momentum inequality in non-vacuum spacetimes'',
  Phys.\ Rev.\ D {\bf 84} (2011) 121503
  [arXiv:1106.3743 [gr-qc]].

\bibitem{Dain:2011}
  S.~Dain, J.~L.~Jaramillo and M.~Reiris,
  ``Area-charge inequality for black holes'',
  Class.\ Quant.\ Grav.\  {\bf 29} (2012) 035013
  [arXiv:1109.5602 [gr-qc]].
  
  \bibitem{Simon:2011}
  W.~Simon,
  ``Bounds on area and charge for marginally trapped surfaces with a cosmological constant'',
  Class.\ Quant.\ Grav.\  {\bf 29} (2012) 062001
  [arXiv:1109.6140 [gr-qc]].
  
  \bibitem{Dain:2011b}
  S.~Dain,
  ``Geometric inequalities for axially symmetric black holes'',
  Class.\ Quant.\ Grav.\  {\bf 29} (2012) 073001
  [arXiv:1111.3615 [gr-qc]].

\bibitem{Clement:2011}
  M.~E.~G.~Clement and J.~L.~Jaramillo,
  ``Black hole Area-Angular momentum-Charge inequality in dynamical non-vacuum spacetimes'',
  Phys.\ Rev.\ D {\bf 86} (2012) 064021
  [arXiv:1111.6248 [gr-qc]].
  
  \bibitem{Jaramillo:2012}
  J.~L.~Jaramillo,
  ``Area inequalities for stable marginally trapped surfaces'',
  arXiv:1201.2054 [gr-qc].
  
\bibitem{Clement:2012}
  M.~E.~G.~Clement, J.~L.~Jaramillo and M.~Reiris,
  ``Proof of the area-angular momentum-charge inequality for axisymmetric black holes'',
  Class.\ Quant.\ Grav.\  {\bf 30} (2013) 065017
  [arXiv:1207.6761 [gr-qc]].
 
 \bibitem{Jaramillo:2012b}
  J.~L.~Jaramillo,
  ``A note on degeneracy, marginal stability and extremality of black hole horizons'', 
  Class.\ Quant.\ Grav.\  {\bf 29} (2012) 177001
  [arXiv:1206.1271 [gr-qc]].
    
  \bibitem{Dain:2013}
  S.~Dain, M.~Khuri, G.~Weinstein and S.~Yamada,
  ``Lower Bounds for the Area of Black Holes in Terms of Mass, Charge, and Angular Momentum'',
  arXiv:1306.4739 [gr-qc].
  
  \bibitem{Dain:GR20}
  Sergio Dain,
  “Geometric inequalities for black holes”,
  GR20, Warsaw, July 2013.
  
  
\bibitem{Visser:2012}
M.~Visser,
``Quantization of area for event and Cauchy horizons of the Kerr-Newman black hole'',
 JHEP {\bf1206} (2012) 023 [arXiv:1204.3138 [gr-qc]].
  
  \bibitem{Faraoni:2012}
  V.~Faraoni and A.~F.~Z.~Moreno,
  ``Are quantization rules for horizon areas universal?'',
  arXiv:1208.3814 [hep-th].
  Accepted for publication in Physical Review D.
  
  
  
  \bibitem{dirty}
   M.~Visser,
  ``Dirty black holes: Thermodynamics and horizon structure'',
  Phys.\ Rev.\ D {\bf 46} (1992) 2445
  [hep-th/9203057].
  
  \bibitem{hernandez-misner}
  W.~C.~Hernandez and C.~W.~Misner,  
  ``Observer time as a coordinate in relativistic spherical hydrodynamics'', 
  Astrophys. J. {\bf143} (1966) 152. [See especially equation (13).]


\bibitem{misner-sharp}
C.~W.~Misner and D.~H.~Sharp, 
``Relativistic equations for adiabatic, spherically symmetric gravitational collapse'', 
Phys. Rev. {\bf136} (1964)  B571.


 \bibitem{symmetric1}
 I.~G.~Macdonald, \emph{Symmetric Functions and Hall Polynomials}, 2nd edition. 
 (Clarendon Press, Oxford, 1995) ISBN 0-19-850450-0 (paperback, 1998).
 
 \bibitem{symmetric2}
Richard P. Stanley. \emph{Enumerative Combinatorics}, Vol. 2. 
(Cambridge University Press, Cambridge, 1995) ISBN 0-521-56069-1. 
  
    
\bibitem{Kerr:1963}
R.~P.~Kerr, 
``Gravitational field of a spinning mass as an example of algebraically special metrics'', 
Physical Review Letters {\bf 11}  (1963) 237--238.  doi:10.1103/PhysRevLett.11.237.

\bibitem{Newman:1965a}
Ezra Newman and Allen Janis, 
``Note on the Kerr Spinning-Particle Metric'', 
Journal of Mathematical Physics {\bf 6} (1965) 915--917.  doi:10.1063/1.1704350.

\bibitem{Newman:1965b}
Ezra Newman, K.~Chinnapared, A.~Exton, A.~Prakash, and R.~Torrence,  
``Metric of a Rotating, Charged Mass'', 
Journal of Mathematical Physics {\bf 6} (1965) 918--919.  doi:10.1063/1.1704351.

\bibitem{Kerr:book}
 D.~L.~Wiltshire, M.~Visser, and S.~M.~Scott (editors),
  \emph{The Kerr spacetime: Rotating black holes in general relativity},
  (Cambridge University Press, 2009).
  
\bibitem{Kerr:survey}
M.~Visser,
  ``The Kerr spacetime: A Brief introduction'',
  arXiv:0706.0622 [gr-qc].
  Published in~\cite{Kerr:book}.


\bibitem{Horowitz:1996}
  G.~T.~Horowitz, J.~M.~Maldacena and A.~Strominger,
  ``Nonextremal black hole microstates and U duality'',
  Phys.\ Lett.\ B {\bf 383} (1996) 151
  [hep-th/9603109].

\end{thebibliography}
\end{document}